\documentclass[aps,pra,epsfigure,twocolumn,showpacs,floatfix]{revtex4}
\usepackage{amsmath}
\usepackage{amstext}
\usepackage{latexsym}
\usepackage{psfig}
\usepackage{graphicx}
\usepackage{amsfonts}

\newcommand{\pro}[2]{\left\vert#1\rangle\langle#2\right\vert}
\newcommand{\ket}[1]{\left\vert#1\right\rangle}
\newcommand{\modul}[1]{\left\vert#1\right\vert}

\newcommand{\bra}[1]{\left\langle#1\right\vert}

\newcommand{\valmed}[1]{\left\langle#1\right\rangle}

\newcommand{\one}{\mbox{$1\hspace{-1.0mm}{\bf l}$}}
\begin{document}

\title{Dynamical entanglement-transfer for quantum information networks}
\author{Mauro Paternostro$^1$, Wonmin Son$^1$,
Myungshik Kim$^1$, Giuseppe Falci$^2$, G. Massimo
Palma$^3$}
\affiliation{$^1$\text{School of Mathematics and
Physics, The Queen's University,
Belfast BT7 1NN, United Kingdom}\\
$^2$\text{ MATIS-INFM \& Dipartimento di Metodologie
Fisiche e Chimiche,
Universita' di Catania,}\\\text{Viale A. Doria 6,
95125 Catania, Italy}\\
$^3$\text{NEST-INFM \& Dipartimento di Tecnologie
dell'Informazione,
Universita' di Milano,}\\\text{Via Bramante 65,26013
Crema, Italy}}
\date{\today}

\begin{abstract}
A key element in the architecture of a quantum information processing network is a reliable physical interface
between fields and qubits. We study a process of entanglement transfer engineering, where two remote qubits
respectively interact with  entangled two-mode continuous variable (CV) field. We quantify the entanglement
induced in the qubit state at the expenses of the loss of entanglement in the CV system. We discuss the range of
mixed entangled states which can be obtained with this set-up. Furthermore, we suggest a protocol to determine
the residual {\it entangling power} of the light fields, inferring, thus, the entanglement left in the field
modes which, after the interaction, are no longer in a Gaussian state. Two different set-ups are proposed: a
cavity-QED system and an interface between superconducting qubits and field modes. We address in details the
practical difficulties inherent in these two proposals, showing that the latter is promising under many aspects.
\end{abstract}
\maketitle
\section{Introduction}

Distributed networks for quantum communication and quantum computation have recently received large attention being a promising architecture for quantum information processing. Efficient sources of entangled continuous variable (CV) states of light are readily available and field modes can act as reliable information carriers. On the other hand, it is more handy to manipulate the information stored in qubits embodied in atomic or solid-state systems. In many of the protocols for quantum information processing designed for distributed networks, entanglement plays a crucial role in establishing an exploitable quantum channel between two distant nodes~\cite{eisert}. This motivates the attempt to study and implement a reliable physical interface between fields and qubits. An interface allows for the transfer of
information from the carrier to the qubit subsystem. In this context, it is interesting to consider the ability of a quantum-correlated two-mode field to induce entanglement in several
different qubit subsystems. 

On the other hand, under a more fundamental point of view, an analysis of the {\it entangling power} of a correlated two-mode state can be a useful tool to indirectly quantify the entanglement left between the modes after the interaction with a given pair of qubits. For a non-Gaussian field state, in particular, there is a lack of objective criteria to determine whether or not entanglement is present between two modes. We show that the capability of the light fields to induce entanglement in two initially separable qubits provides a test for the inseparability of the fields.

The implementation of such a physical interface, thus, opens a way to investigate the exchange of
entanglement between systems defined in heterodimensional Hilbert spaces~\cite{wonmin,kraus,mauro} and to the
{\it entanglement transfer} processes, where entanglement increases in one system at the expense of the loss of the
other one, through their interaction.

In this work, we study a two-qubit system interacting with a quantum-correlated two-mode squeezed state of light
through bi-local resonant interactions. The system we analyze allows to engineer entanglement transfer,
generating qubit states with a controllable amount of entanglement. We show that it is actually possible to
explore a large part of the space of the entangled mixed states (EMS) of two qubits, including the class of boundary states having the maximum amount of entanglement for a given value of the purity of the state~\cite{munro1}. The generation of arbitrary and controllable entangled mixed states is relevant to investigate about the role that entanglement and purity have for the tasks of Quantum Information Processing (QIP). It has been proved, for example, that while entanglement is a fundamental requirement to efficiently perform the Shor's factorization, purity is not~\cite{parkerplenio}. Moreover, in some cases, efficient QIP can be performed even relaxing the requirements, in terms of quantum correlations and purity, of a quantum channel. For example, a neat threshold exists for the allowed mixedness of an entangled mixed state to be usefully used in a teleportation protocol~\cite{bosevedral}.

The interest in the generation of EMS and in quantifying the entangling power of a given two-mode entangler leads us to look for practical implementations of the model we propose here. We consider two different set-ups to implement our model, namely a cavity-quantum electrodynamics (cavity-QED) system and Josephson charge qubits interfaced to field modes. We describe the two systems in some  details and compare their performances. In the first proposal we consider two optical cavities, each interacting with one of the modes of a two-mode squeezed state of light respectively. After the squeezed state feeds the cavities, two two-level atoms cross their respective cavities.
However, this set-up faces important practical difficulties. The cavity-photon lifetime turns out to be
comparable to the operation times of our set-up, a feature that is typical of the {\it weak coupling regime} of
the atom-field interaction. On the other hand, the advantages of a much stronger coupling~\cite{buisson,francescopino} 
can be exploited in the second scenario we propose, where the qubits and the cavity are implemented by 
superconducting devices integrated on the same chip. In this case, flying qubits are no more required so that the control
over the interaction times is easier. Moreover, in this scheme the qubit parameters can be independently modulated via external electric and magnetic fields~\cite{Schon}. 
The interaction can be controlled by tuning the qubit on/off resonance with a cavity mode~\cite{francescopino,mauro}.
A promising design of integrated superconducting qubit and 
cavity has been recently
proposed~\cite{schelkopf} and the first experiments have
already demonstrated a quality factor $Q > 10^4$ for the 
cavity~\cite{schelkopf-priv}, which is enough to implement the
protocols discussed in refs.~\cite{francescopino,mauro}.


The paper is organized as follows. In Section~\ref{approccio} we describe the general scheme proposed here,
introduce the space of the entangled mixed states of two qubits and show that our model allows for an extensive
exploration of such a space. In Section~\ref{estensione}, we infer the correlations left between two field modes
after the interaction. We propose a scheme that, extending the interaction model to more than a single qubit
pair, allows to get some additional information on the residual entanglement capabilities of the field,
bypassing the lack of a necessary and sufficient criterion for the entanglement of a non-Gaussian state. In
Section~\ref{implementazioni}, we describe the proposed implementations of our system. For a cavity-QED system,
we discuss the effect of spectator atoms and of the cooperativity parameter~\cite{turchettekimble}. This kind of
difficulties are no more present
 when Cooper pair boxes~\cite{Schon} integrated in
high-quality transmission lines are used.


\section{The model of entanglement generation}
\label{approccio}

\begin{figure} [ht]
\centerline{\psfig{figure=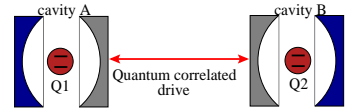,width=4.4cm,height=1.4cm}}
\caption{Two remote cavities are driven by a quantum correlated two-mode field. We explicitly consider the case of a two-mode squeezed driving field coupled to each cavity via a leaky mirror and preparing a quantum-correlated two-mode state of the cavity fields. After the preparation of the cavity field, local interactions with two generic (flying or static) qubits are arranged.}
\label{sistema0}
\end{figure}

We introduce the system we consider, based on the interaction of a pair of remote qubits with local environments that, for definiteness, we model as single bosonic modes, respectively. While the physical system that embodies the qubits will not be specified until necessary, this schematization perfectly matches the situation in which the two qubits are coupled  to the field modes of two cavities. This assumption does not limit the generality of our approach allowing, on the other hand, to cover many different physical systems that are promising for the purposes of QIP~\cite{massimolibro}. We will consider explicitly this case, from now on. We are interested in a situation in which the field modes of the two cavities exhibit non-classical correlations and local interactions of each qubit with the respective cavity mode are arranged. A way in which this can be achieved is sketched in Fig.~\ref{sistema0}.

In details: two remote but identical cavities are initially prepared in their vacuum states. One mirror of each
cavity is assumed to be perfect while the other has non-zero transmittivity. The leaky mirror is coupled
to one mode of an external two-mode squeezed state with coupling strength $\kappa$. The squeezed state of the two external modes $a$ and $b$ is represented by $\ket{S}_{ab}=(\cosh{r})^{-1}\sum^{\infty}_{n,0}(\tanh{r})^{n}\ket{n}_a\ket{n}_{b}$, where $\left\{\ket{n}\right\}_{a,b}$ are the Fock bases for the field modes, and $r$ is the squeezing parameter. If the coherence time of the driving field is shorter than $\kappa^{-1}$, we can treat the cavity mirror as a beam splitter 
continuously fed by squeezed fields~\cite{myungimoto}. The state of the cavity modes $A$ and $B$ evolves due to the coupling and is described by the reduced density matrix~\cite{wonmin}
\begin{equation}
\label{statocavita'}
\begin{aligned}
&\rho_{AB}=\mbox{Tr}_{ab}\left(\hat{{\cal
B}}_{Aa}(\theta)\hat{{\cal B}}_{Bb}
(\theta)\rho_{abAB}(0)\hat{{\cal
B}}^{\dag}_{Aa}(\theta)\hat{{\cal B}}^{\dag}_{Bb}
(\theta)\right)\\
&=\hskip-3pt \sum^{\infty}_{n,m=0}\hskip-3pt 
\sum^{\min(n,m)}_{k,l=0} \hskip-3pt
{\cal A}^{nm}_{kl}
\ket{n-k}_{A}\!\bra{m-k}\otimes\ket{n-l}_{B}\!\bra{m-l}.
\end{aligned}
\end{equation}
Here, $\hat{B}_{Aa}(\theta)=\exp\left[{\theta}(\hat{A}\hat{a}^{\dag}-\hat{A}^{\dag}\hat{a})\right]$ is the beam splitter operator~\cite{barnett} with its reflection coefficient $\sin\theta$ (basically determined by $\kappa$~\cite{myungimoto}), $\hat{a}$ ($\hat{a}^{\dagger}$) is the annihilation (creation) operator for the external field mode $a$ (analogously for field mode $b$) while $\hat{A}$ ($\hat{A}^\dag$) and $\hat{B}$ ($\hat{B}^\dag$) are the bosonic operators of the cavity fields. We have defined ${\cal A}^{nm}_{kl}=\chi_{nm}G^{nm}_{kl}$, with $\chi_{nm}=(\tanh{r})^{n+m}/{(\cosh{r})}^2$ and $G^{nm}_{kl}=\prod^{m}_{\alpha=n}\prod^{l}_{\beta=k} \sqrt{\alpha{!}/[\beta{!}(\alpha-\beta)!]} (\cos\theta)^{\beta}(\sin\theta)^{\alpha-\beta}$. When the reflected part of the external field is minimized, a two-mode squeezed state  quickly builds up inside the cavities. Note that a two-mode Gaussian field is built into the cavity, regardless of the coupling strength. This model is exactly the same as the initial field interacting with a zero-temperature reservoir~\cite{myungimoto} and it is known that entanglement can always be found during the evolution of an initial two-mode squeezed field in the zero-temperature vacuum~\cite{duan}.

After the cavity field is prepared, the interaction with a pair of qubits with logical states $\ket{0}_{i}$ and $\ket{1}_{i}$ ($i=1,2$) begins. Here, the interaction model is assumed to be of the
resonant Jaynes-Cummings type~\cite{jc}. The interaction Hamiltonian for the cavity A is
\begin{equation}
\hat{H}_{A1}=\hbar\Omega\left(\hat{A}\ket{1}_1\!\bra{0}+
\hat{A}^\dag\ket{0}_1\!\bra{1}\right), \label{ham}
\end{equation}
where $\Omega$ is the atom-field coupling constant. An analogous  Hamiltonian describes the interaction between
the second qubit and cavity B. Before entering the detail of protocols, we point out 
that the Hamiltonian we discuss can be also implemented by
superconducting devices~\cite{buisson,francescopino,schelkopf}.

The interaction between cavity modes and  qubits gives rise to an unitary evolution 
$\hat{U}_{A1}(t) \otimes \hat{U}_{B2}(t)$ of the whole system, where $\hat{U}_{A1}(t)=\exp(-i\hat{H}_{A1}t/ \hbar)$. The effective evolution of the two qubits, on the other hand, is non-unitary and is described by the 
reduced density matrix $\rho_{12}(t)$ obtained by tracing out the cavity fields as $\rho_{12}(t)=\mbox{Tr}_{AB}\left(\hat{U}_{A1}(t) \otimes \hat{U}_{B2}(t)\rho_{12}(0)\otimes\rho_{AB}\hat{U}^{\dag}_{A1}(t) \otimes \hat{U}^{\dag}_{B2}(t)\right)$. Here, we consider the initial state $\rho_{12}(0)=\ket{00}_{12}\!\bra{00}$~\cite{commentoM} and define the rescaled time of this first interaction $\tau_{1}=\Omega{t}$. 
In the basis $\left\{\ket{11},\ket{10},\ket{01},\ket{00}\right\}_{12}$, $\rho_{12}(\tau_{1})$ takes the
form
\begin{equation}
\label{matricedensita} \rho_{12}(\tau_{1})=
\begin{pmatrix}
A&0&0&-D\\
0&B&0&0\\
0&0&C&0\\
-D&0&0&F
\end{pmatrix}
,
\end{equation}
with $F=1-A-B-C$,
\begin{equation}
\begin{split}
&A=\sum^{\infty}_{n=0}\sum^{n}_{k,l=0}{\cal
A}^{nn}_{kl}\sin^{2}(\tau_{1}\sqrt{n-k})\sin^{2}(\tau_{1}\sqrt{n-l}),\\
&B=\sum^{\infty}_{n=0}\sum^{n}_{k,l=0}{\cal
A}^{nn}_{kl}\sin^{2}(\tau_{1}\sqrt{n-k})\cos^{2}(\tau_{1}\sqrt{n-l}),\\
&C=\sum^{\infty}_{n=0}\sum^{n}_{k,l=0}{\cal
A}^{nn}_{kl}\cos^{2}(\tau_{1}\sqrt{n-k})\sin^{2}(\tau_{1}\sqrt{n-l}),\\
\end{split}
\label{diagonal}
\end{equation}
and the off-diagonal component given by
\begin{equation}
\begin{split}
D&=\sum^{\infty}_{n=0}\sum^{n}_{k,l=0}{\cal
A}^{nn+1}_{kl}\sin(\tau_{1}\sqrt{n-k+1})\\
&\times\cos(\tau_{1}\sqrt{n-k})\sin(\tau_{1}\sqrt{n-l+1})\cos(\tau_{1}\sqrt{n-l}).\\
\end{split}
\label{off-diagonal}
\end{equation}

In order to quantify the quantum correlations between
the qubits after the interaction with the cavity modes, we use
the negativity of the partially transposed density
operator (NPT), which is a necessary and sufficient condition for
entanglement of any bipartite qubit system~\cite{zyczkowski}. The
entanglement measure based on NPT is defined as ${\cal
E}_{NPT}=-2\lambda^{-}$, where $\lambda^{-}$ is the
negative eigenvalue of the partially transposed density matrix
$\rho^{PT}_{12}$ (here with respect to qubit
$2$)~\cite{leekim}.
In our case, $\lambda^{-}$ does not depend on the
populations of states $\ket{11}_{12},\,\ket{00}_{12}$. Explicitly
\begin{equation}
\label{negativeeigenvalue}
\lambda^{-}=\frac{1}{2}\left\{{B}+C-\sqrt{4D^2+{\left(B-C\right)^2}}\right\}.
\end{equation}

\begin{figure}[b]
\centerline{\psfig{figure=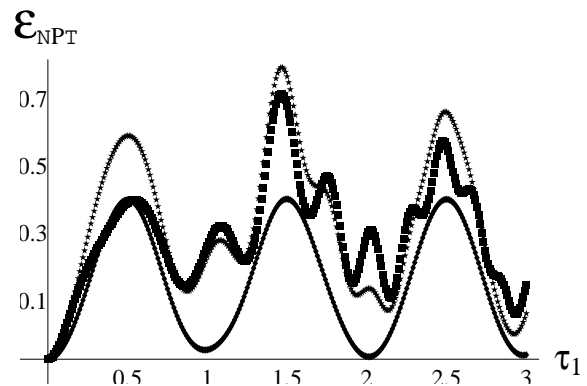,width=8.0cm,height=5.0cm}}
\caption {${\cal E}_{NPT}$ versus the rescaled
interaction time $\tau$ (in units of $\pi$).
The squeezing parameter of the field is $r=0.26$ (full line),
$r=0.86$ (stars), $r=1.2$ (full squares). These values offer a trade-off between the 
visibility of each curve and the insight given in the dynamics of the entanglement. We take
$\sin^2 \theta = 0.1$.}
\label{entanglementtempo}
\end{figure}

In Fig.~\ref{entanglementtempo} we plot ${\cal E}_{NPT}$ versus $\tau_{1}$ for a reflectivity of the cavity $\sin^2\theta=0.1$ and three different values of
the squeezing parameter $r$. The entanglement is peaked at $\tau_{1}=(2q+1)\pi/2$ ($q$ integer). For small values of $r$, the two-mode squeezed state can be approximated by $\ket{00}_{ab}+{r}\ket{11}_{ab}$ so that the qubit-field interaction results in $\ket{00}_{12}-{r}\ket{11}_{12}$ at $\tau_{1}=(2q+1)\pi/2$.
However, as squeezing is
increased, the Rabi oscillations
become more complicated due to the importance of terms 
relative to higher photon numbers
in eqs. (\ref{diagonal}) and (\ref{off-diagonal}). 
This reduces the qubit entanglement and makes ${\cal E}_{NPT}$ a non-monotonous function of the squeezing. This is explicitly shown in Fig.~\ref{entanglementtempo}, for $r=1.2$ (full squares). In this case, the entanglement function is always below the curve relative to the smaller $r=0.86$ (stars) that allows for the maximum ${\cal E}_{NPT}$ at $\tau_{1}\approx{3\pi/2}$.

\begin{figure} [ht]
\centerline{\psfig{figure=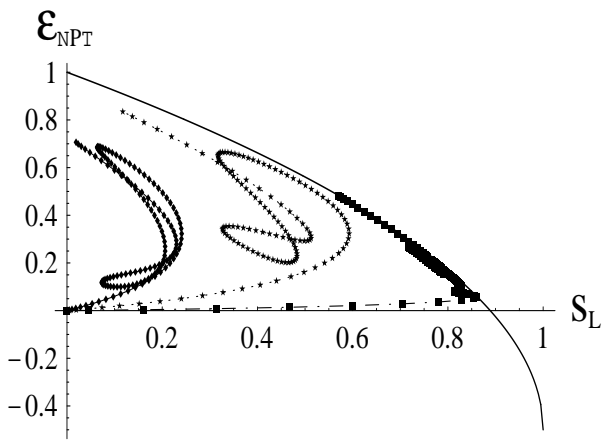,width=8.0cm,height=6.0cm}}
\caption{Navigation in the plane of the entangled
mixed states of two qubits. We used 
$\sin^2 \theta=0.1$. The measures for the entanglement and
purity are negativity of partial transposition ${\cal E}_{NPT}$
and linearized entropy $S_{L}$. The curvilinear abscissa is 
$\tau_{1}\in[0,3\pi/2]$. We show the cases with
$r=0.46$ (rhombuses), $r=0.86$ (stars) and $r=2$ (filled
squares). The solid line is the MEMS boundary.}
\label{navigation1}
\end{figure}

Another useful quantity to characterize the state of the qubits is the degree of purity of $\rho_{12}$. We have
studied the behavior of the linearized entropy
$S_{L}(\rho_{12})=(4/3)\left[1-Tr\left(\rho^2_{12}\right)\right]$, which gives a measure of the degree of mixedness for the state described by $\rho_{12}$~\cite{munro1}. $S_{L}(\rho_{12})$ ranges from $0$ (pure states) to $1$ (maximally mixed states). The linearized entropy can be studied as a function of $\tau_{1}$ and for different values of $r$. We have combined the temporal behavior of entanglement and linear entropy onto an {\it entanglement-purity} plane as shown in Fig.~\ref{navigation1}.
In this plot, each curve is relative to a specific
value of the squeezing parameter, while each point along a curve
gives the entanglement and purity of the corresponding state at
a specific interaction time (so that $\tau_{1}$ represents a curvilinear abscissa, in this plot). The solid line is the upper bound to
the region occupied by physically achievable mixed
states for a given degree of entanglement ${\cal E}_{NPT}\ge{0}$.
They are known as {\it maximally entangled mixed states} (MEMS)~\cite{munro1}.
A parameterization of MEMS is critically dependent on the
chosen measures of entanglement and purity. If the negativity of partial
transposition is taken, there are two classes of one-parameter states
belonging to MEMS and giving the
same boundary. The first class corresponds to the
family of Werner states
$\rho_{W}=p\pro{\phi^+}{\phi^+}+(1-p)\one_{2\times2}$,
with $\ket{\phi^{+}}=(1/\sqrt{2})\left(\ket{00}+\ket{11}\right)$
and $0\le{p}\le1$. The other class can be 
resented by
\begin{equation}
\label{memsmunro2}
\rho_{MEMS}=
\begin{pmatrix}
\frac{1+\sqrt{1+3p^2}}{6}&0&0&\frac{p}{2}\\
0&\frac{2-\sqrt{1+3p^2}}{3}&0&0\\
0&0&0&0\\
\frac{p}{2}&0&0&\frac{1+\sqrt{1+3p^2}}{6}
\end{pmatrix}.
\end{equation}
Thus, the Werner states represent a frontier for the
achievable amount of entanglement that a given mixed state can
have. This feature is unique for the choice of NPT as the measure
of entanglement~\cite{munro1}.

Several interesting points can be addressed closely analysing Fig.~\ref{navigation1}. First of all, the
interaction model described here is able to produce entangled, nearly pure, state of two qubits (see, for example, the last point on the curve relative to $r=0.46$) that can be used to test protocols for QIP~\cite{bosevedral}. As the value of  $r$ increases, the states of the qubits get closer to the frontier curve of MEMS, even if the corresponding states are weakly entangled for longer periods of time.
As was pointed out, this is due to the contribution of highly excited photon number states, in the Schmidt
decomposition of a two-mode squeezed state $\ket{S}$, for larger values of $r$~\cite{wonmin}. Our theoretical model is flexible enough to generate arbitrary entangled mixed states of two qubits up to the boundary-class of MEMS. Recently, theoretical and experimental efforts  have been performed in the generation of these states using different mechanisms~\cite{parkinskwiat}. In our scheme, the local interactions with the two-mode entangler induces time-dependent quantum correlations between the qubits. The mixedness of the two-level systems state  is due to their entanglement with the cavity modes. We will see that, in fact, the entanglement between the qubits is set at the expenses of the correlations between the field modes. 


\section{Entanglement dynamics for cavity fields}
\label{estensione}

In the previous section, we studied the amount of entanglement generated in a bipartite two-level system by the
interaction with a two-mode squeezed state. Now, we turn our attention to the entanglement left between the
cavity fields. After the interaction, the cavity fields state is given by
\begin{eqnarray}
\hat{\rho}_{AB}(t)=\mbox{Tr}_{12}\left(\hat{U}_{AB12}(t){\rho}_{AB}(0)
\otimes{\rho}_{12}(0)
\hat{U}_{AB12}^{\dagger}(t)\right),
\end{eqnarray}
where $\hat{U}_{AB12}(t)=\hat{U}_{A1}(t)\otimes\hat{U}_{B2}(t)$.
It is not difficult to see that the cavity field is no
more Gaussian after the interaction.

Differently from the qubit case, how to quantify the quantum correlations of a CV state is only partially known. For the case of a
Gaussian state, NPT can be used as a separability criterion as well as a measure of
entanglement~\cite{Simon,DGCZ}. In this case, the NPT condition for separability is equivalent to the violation
of the uncertainty principle by a partially transposed density operator~\cite{Simon}. To study the uncertainty
principle, it is useful to consider the vector of the field quadratures
$\hat{\xi}=(\hat{q}_1,\hat{p}_1,\hat{q}_2,\hat{p}_2)^{T}$, where
$\hat{q}_{1}=\left(\hat{A}+\hat{A}^{\dag}\right)/\sqrt{2}$,
$\hat{p}_{1}=-i\left(\hat{A}-\hat{A}^{\dag}\right)/\sqrt{2}$ and $\hat{q}_{2},\,\hat{p}_{2}$ are analogously
defined in terms of $\hat{B}$ and $\hat{B}^{\dag}$. The field quadratures satisfy the commutation relations
$[\hat{\xi}_{\alpha},\hat{\xi}_{\beta}]=i\Omega_{\alpha\beta}$, where $\Omega_{\alpha\beta}$ are the elements of
the $4\times{4}$ matrix ${\bf \Omega}=\oplus^{2}{\bf J}$, with ${\bf J}=\begin{pmatrix}0&1\\-1&0\end{pmatrix}$.
Some of the statistical properties of a two-mode CV state can be inferred from the $4\times 4$ {\it covariance
matrix} $\bf V$, defined by
$V_{\alpha\beta}=\langle\Delta\hat{\xi}_{\alpha}\Delta\hat{\xi}_{\beta}+\Delta\hat{\xi}_{\beta}\Delta\hat{\xi}_{\alpha}\rangle/2$,
with ${\Delta\hat{\xi}}_{\alpha}=\hat{\xi}_{\alpha}-\valmed{\hat{\xi}_{\alpha}}$ and the expectation values
evaluated over the state of the light fields. In terms of ${\bf V}$, the uncertainty relation for the field
quadratures takes the form ${\bf V}+i{\bf \Omega}\ge0$. Using the block representation ${\bf V}=\begin{pmatrix}
{\bf A} & {\bf C}\\
{\bf C}^{T}& {\bf B}
\end{pmatrix}$,
with ${\bf A, B}$ and ${\bf C}$ $2\times{2}$ matrices,
the uncertainty relation can be restated as~\cite{Simon}
\begin{equation}
\label{uncertain}
\begin{split}
\Delta&\equiv(\det {\bf A})(\det {\bf B})+
\left({1}-\det {\bf C}\right)^2-\mbox{Tr}({\bf
\tilde{A}}{\bf \tilde{C}}{\bf \tilde{B}}{\bf
\tilde{C}}^{T})\\
&-(\det{\bf A} + \det {\bf B})\ge{0}.
\end{split}
\end{equation}
with ${\bf \tilde{A}}={\bf AJ}$, ${\bf \tilde{B}}={\bf
BJ}$ and ${\bf \tilde{C}}={\bf CJ}$. Each term of
$\Delta$ is invariant under local linear canonical
transformations. Partial transposition is equivalent to a mirror reflection (that performs the transformation $\hat{p}_{i}\rightarrow-\hat{p}_{i}$) in the phase-space. This changes the sign of $\det{\bf C}$, leaving the other terms unaffected~\cite{Simon}. Thus the uncertainty $\Delta_{NS}$ for the partially transposed density matrix is obtained replacing ${\bf C}$ by $\modul{\det{\bf C}}$ in Eq.~(\ref{uncertain}). In our study, the function $\Delta_{NS}$ depends on the squeezing parameter $r$ and the interaction time $\tau_{1}$. For a Gaussian state, $\Delta_{NS}<0$ is a necessary and sufficient condition for entanglement.

For non-Gaussian states the situation is more complicated. 
The violation of the Heisenberg uncertainty relation by 
$\rho^{PT}_{AB}$ is only a sufficient condition for
non-separability of the state. Nevertheless, as far as the authors are concerned, the uncertainty criterion is one of the most successful conditions to test entanglement of even a non-Gaussian state. We challenge this condition in the following.

\begin{figure} [ht]
\centerline{\psfig{figure=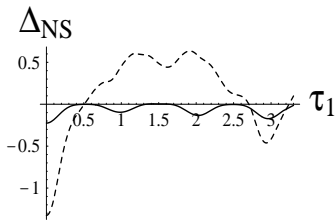,width=4.5cm,height=3.0cm}}
\caption{This plot shows the dynamics of the
uncertainty relation with partially transposed density matrix. Negative
values show the violation of uncertainty or, equivalently, the
non-separability of the cavity field modes $A$ and $B$ depending on the
atom-field interaction time. We set $(\sin\theta)^2=0.1$ with
$r=0.46$ (solid line) and $r=0.86$ (dotted line).}
\label{nonsep}
\end{figure}

In Fig.~\ref{nonsep}, $\Delta_{NS}$ for the cavity
fields after the interaction with the qubits is
plotted against $\tau_{1}$ for two significant values of
the squeezing parameter. As done before, we assume the qubits initially prepared in their
ground states $\ket{00}_{12}$. Interestingly, $\Delta_{NS}$ is negative for
most of the interaction time, when $r=0.46$ (solid line). This
means that the cavity field modes are in an entangled state even after the qubit-field
interaction. The dynamics of $\Delta_{NS}$ 
is opposite to the qubit
entanglement as it is apparent comparing 
Figs.~\ref{entanglementtempo} and~\ref{nonsep}. This
behavior is consistent with the idea that the entanglement in the field is
transferred to the qubits. 
On the other hand, for $r=0.86$
the entanglement induced between the
qubits does not really match to the behavior of $\Delta_{NS}$
(dashed line). 

More informations about the entanglement of the field modes
 can be obtained testing their entangling power, that is their ability
to set entanglement in an additional pair of initially separable
qubits. In this case, if the cavity fields are not 
entangled there is no way to transfer correlations to two additional 
qubits. A nonzero entangling power is, thus, a condition for the fields 
states to be inseparable.

\begin{figure} [ht]
\centerline{\psfig{figure=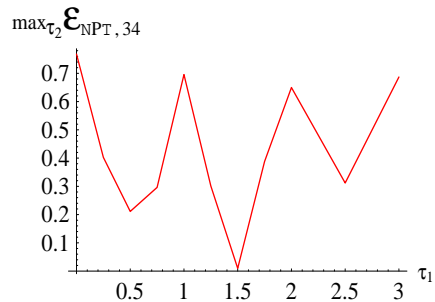,width=6.0cm,height=4cm}}
\caption{This plot shows the entanglement
between qubits $3$ and $4$, maximized with respect to the interaction time 
$\tau_{2}$, as a function of $\tau_{1}$ (units of $\pi$). Because of the limit in computational power, the maximum of ${\cal E}_{NPT,34}$ is calculated for the cases of the first pair interaction times $\tau_{1}=n\pi/4$ (n=0,1,..,8).}
\label{2pair}
\end{figure}

We let a second pair of qubits (indexed by 3 and 4)
interact with the respective 
cavities for a time $\tau_{2}$. 
In general, the degree of entanglement 
${\cal E}_{NPT,34}$ between the qubits 
depends also upon $\tau_{1}$, 
the interaction time with the first pair. 
In Fig.~\ref{2pair} we plot
$\max_{\tau_{2}}{\cal E}_{NPT,34}(\tau_{1},\tau_{2})$, 
that is the maximum achievable entanglement ${\cal E}_{NPT,34}$
for a given interaction time $\tau_{1}$.
We consider $r=0.86$ and all the qubits initially in
their ground state. Obviously, if the
first pair of qubits does not interact with the
cavities ($\tau_{1}=0$), the entanglement settled in the second
pair is the maximum achievable. Comparing 
Figs.~\ref{entanglementtempo} and~\ref{2pair}, we note that 
the second pair may be entangled more as the first pair is 
entangled less and vice-versa. Notice that at $\tau\simeq{1.5}$, i.e. when 
the entanglement of the first pair is maximal 
(see Fig.~\ref{entanglementtempo}), we find that 
${\cal E}_{NPT,34}$ is nonzero, showing that 
even in this case the entanglement
capability of the cavity fields is not exhausted by the 
first interaction. Thus the field modes in the
cavities are still quantum-mechanically correlated and able
to entangle the two qubits by the local interactions, despite
the fact that $\Delta_{NS}$ is positive.
In this example, the entanglement capability is a more 
powerful test for the entanglement of the cavity fields.

\section{Two physical implementations}
\label{implementazioni}

The generation of two-qubit quantum states up to the MEMS boundary and the interest in inferring the entangling capabilities of a non-Gaussian CV state motivate the search for an implementation of the model studied so far. In this Section we describe two different proposals for a set-up: a cavity-QED scheme and an interface between superconducting charge qubits and cavity field modes. This latter, in particular, offers some intriguing perspectives in terms of coherence times and control of the qubits.

\subsection{Cavity-QED system}
\label{cavita}

\begin{figure} [ht]
\centerline{\psfig{figure=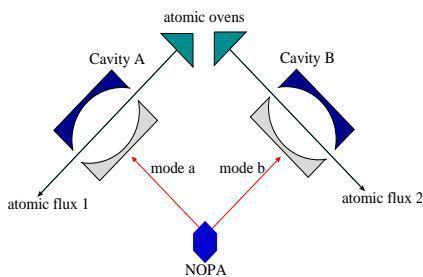,width=5.5cm,height=3.5cm}}
\caption{Two remote optical cavities are fed by the
two-mode squeezed state generated by a
nondegenerate-parametric-amplifier
(NOPA). The external fields interact with two cavity
modes via direct coupling to the leaky mirror of each
cavity (the other being perfect). A low-density flux of
two-level atoms, initially prepared in their ground states $\ket{g}$,
passes through each cavity and interacts with the relevant
cavity mode.}
\label{sistema}
\end{figure}

The first physical set-up we analyze is sketched in Fig.~\ref{sistema}. As outlined in Section~\ref{approccio},
two remote one-sided optical cavities are initially prepared in the vacuum state. A two-mode squeezed state is
coupled to each cavity via the leaky mirrors. In what follows, we assume a {\it broad-band} external source. We
call $\Delta\omega_{ext}$ the band-width of the dyriving source. To consider the external field as an infinite
band-width drive and use the (simple) analytical expressions valid for a broad-band squeezed state, the
condition $\kappa\ll\Delta\omega_{ext}$ has to be fulfilled. However, it is typically
$\Delta\omega_{ext}\lesssim{\kappa}$
~\cite{turchettekimble}. This means that to compare
any theoretical prediction to the results of a real
experiment, a more involved finite band-width approach
must be used~\cite{turchettekimble}. This goes beyond the
tasks of this work and we will assume the broad-band
condition for our theoretical investigation. We assume the cavity fields build up as a two-mode squeezed state before the interactions with the qubits start.

The qubits are here embodied by two flying two-level atoms of their ground and excited states $\ket{g}_{i},\,\ket{e}_{i}$ ($i=1,2$) that pass
through the respective cavities.

The behavior of a one-sided cavity, with respect to an
external driving field, is influenced by the presence
of an atomic medium.
A two-level atom in a resonator
can be seen as an intra-cavity lossy medium with a loss parameter
proportional to the {\it atomic cooperativity}
$C=\Omega^2/\kappa\Gamma$, where $\Gamma$ is the
spontaneous emission rate from
$\ket{e}$. For large $C$, the intra-cavity losses are
so large that, eventually, no squeezed state builds up in the
resonators~\cite{turchettekimble}. 
Realistic parameters, within the state of the art, are $(\kappa,\Omega,\Gamma)/2\pi=(80,40,4)\,MHz$. While the
interaction of a cavity mode with the respective flying atom arises naturally in the form of a Jaynes-Cummings
Hamiltonian, an important issue to take into account is the strength of this interaction. The optimal condition
would be, obviously, the {\it strong coupling regime} $\Omega\gg\kappa,\Gamma$, where the coherent evolution of
the atom-cavity system is faster than the decoherence mechanisms (dissipation, dephasing) due to the cavity decay
and to the atomic spontaneous emission~\cite{commento}. In general strong coupling with flying atoms is a hard
task to achieve, in optical systems~\cite{commento2}. Furthermore, with the above values for $\kappa$, suitable
for the external field to be injected, the field coherence times are within the range of $10\,nsec$. On the
other hand, for cavity field waist of $\sim10\,\mu{m}$ and an atomic velocity $\sim300\,m/sec$ (that is an
optimistic value) we get transit times in the range of $10\,nsec$, comparable to $\kappa^{-1}$. The dissipative
effect due to the cavity decay is, thus, not negligible and experimental efforts are required to give a
practical implementation of our scheme in this set-up. This has some implications for our method to infer the
entangling capabilities of the fields by detecting the correlations between the atoms of a second pair.
Intuitively, the time elapsing between the passages of two consecutive atoms has to be shorter than the
life-time of a photon in an optical cavity. Otherwise, the decay of the cavity fields will destroy the quantum
correlations between the field modes. The effects of the cavity decay can be described by introducing a
dissipative term $\hat{H}_{loss}=-i\hbar\kappa\sum_{\alpha}\hat{\alpha}^{\dag}\hat{\alpha}$ ($\alpha=A,B$) in
the system Hamiltonian. This term gives an exponential decay (with rate $\kappa$) of the
probability to find the cavity mode $\alpha$ in a Fock state with $p$ photons. We call $\bar{t}$ the time elapsing
between the passage of the first and the second pair of qubits. It turns out that
, for $\bar{t}$ equal to an atomic transit time (so that the first pair of atoms has surely left the region of
interaction with the cavities), ${\cal E}_{NPT,34}$ is $50\%$ less than what we get in the ideal conditions. A way to minimize the elapsing time is to have the simultaneous presence of at least two qubits in
the same cavity. For the sake of definiteness, we suppose the intensity of each cavity field to have a Gaussian
radial profile centred at the cavity axis. We assume that, while the first atom is interacting with the cavity
field, the second is at the border of the region of interaction and is weakly coupled to the field. We thus have
a {\it spectator atom} inside the cavity~\cite{turchettekimble}. Usually, the spectator atoms give rise to
additional loss mechanisms that become important once the density of the spectators is such that their {\it
collective} coupling to the cavity mode is comparable to $\Omega$. A way to reduce these losses would be, then,
to control a very low-density atomic beam.

Having fixed the interaction time (that can be finely controlled),
a possible source of errors, in our model, is represented by the
mismatched injection of the atoms in the cavity.
The distribution of the atomic velocities is
thermal and it is, in general, hard to arrange the simultaneous entrance of
two atoms in two remote cavities. Obviously, a certain control
is possible by means of atomic cooling techniques. However, a quantitative analysis of the effect of a mismatched triggering of the qubit-cavity
mode interactions shows that the qubit-qubit
entanglement is flattened to zero for times short compared to the interaction time $\tau_{1}$. The long-time behavior of ${\cal E}_{NPT,12}$, then, follows the pattern expected for a perfectly injected pair of qubits. In general, for a delay $\delta{t}$ between the entrance of $1$ and $2$ much smaller than $\Omega^{-1}$, this effect on the generation of a pair of entangled atoms is negligible. Obviously, a complete description of the effect
of a mismatched atomic injection is given averaging
the entanglement over $\delta{t}$. Such a detailed
analysis is not necessary whenever we are able to keep the delay
times within the condition $\delta{t}\ll\Omega^{-1}$. However, the control of  $\delta{t}$ is {\it per se} a hard task. Some of the problems faced by this optical cavity-QED implementation can be solved in a scenario in which microwave cavities, interacting with long-lived Rydberg atoms, are considered. In this case, however, there is still the need for a fine control of the transit times and the simultaneous entrance of the atoms inside the cavities. 

\subsection{Josephson qubits in a superconducting
transmission line}
\label{giunzioni}
We now consider a second class of implementations
that combines some of
the features of a microwave cavity-QED 
with the characteristics of a superconducting
nano-circuit.
The first advantage of a solid state device combined with quantum optics is the 
possibility of achieving a strong qubit-field mode
Jaynes-Cummings type interaction~\cite{buisson}. This is basically due to the fact that
 the cavity field, in this case, is coupled to the charge of the qubit rather than its 
dipole. Various solid state implementations of the 
cavity have been proposed, ranging from large Josephson
junctions~\cite{buisson,francescopino} to superconducting films with 
large kinetic inductance~\cite{buisson} to microstrip 
resonators~\cite{schelkopf}. Several 
operational schemes for quantum protocols have been suggested, in this contexts~\cite{super-cavity-th,francescopino,mauro}. 
Coupling between a superconducting qubit in the charge 
regime~\cite{Schon,Nakamura} 
and a classical Josephson junction 
has been implemented 
by the Saclay group~\cite{vion}. Decoherence 
of a qubit coupled with a quantum oscillator in the solid 
state has been studied in ref.~\cite{francescopino}.
This analysis has revealed the existence of 
optimal operating conditions where decoherence is essentially
due to spontaneous non-radiative decay of the qubit and 
leakage from the computational space of the cavity modes. 
This latter source of decoherence is
minimized if the cavity is implemented by an 
integrated high-$Q$
microstrip resonator~\cite{schelkopf} so, in what follows, we will focus on 
this particular implementation. 

In the proposal of ref.~\cite{schelkopf}, a Superconducting-Quantum-Interference-Device (SQUID) qubit~\cite{Schon} 
is placed between the planes of the
resonator and the whole device is fabricated using 
nanolithographic techniques.
The geometry of
the resonator is such that a single mode of frequency
$\sim{10}\,GHz$ can be accommodated.
Microstrip transmission-lines having a $Q$ factor of
about $10^4$ (corresponding to
photon life-time of $0.1\mu{sec}$) have been already built  and the effect of coupling
with a qubit implemented by a SQUID has been 
demonstrated~\cite{schelkopf-priv}. Realistic estimates of the
qubit life-time are in the range of some $\mu{sec}$ 
and the whole situation is reminding of a microwave cavity-QED
set-up where the internal losses due to the cavity
decay are very low. The source of quantum radiation can be built up using nondegenerate Josephson parametric amplifiers (in a distributed configuration, to limit the effects of gain-increasing noise). Microwave squeezed radiation (in a range of frequency of $\sim{10}\,GHz$) has been demonstrated and the source impedance-matched to superconducting lines used to transport the signals~\cite{yurke}. 

\begin{figure} [ht]
({\bf a})\hskip3cm({\bf b})
\centerline{\psfig{figure=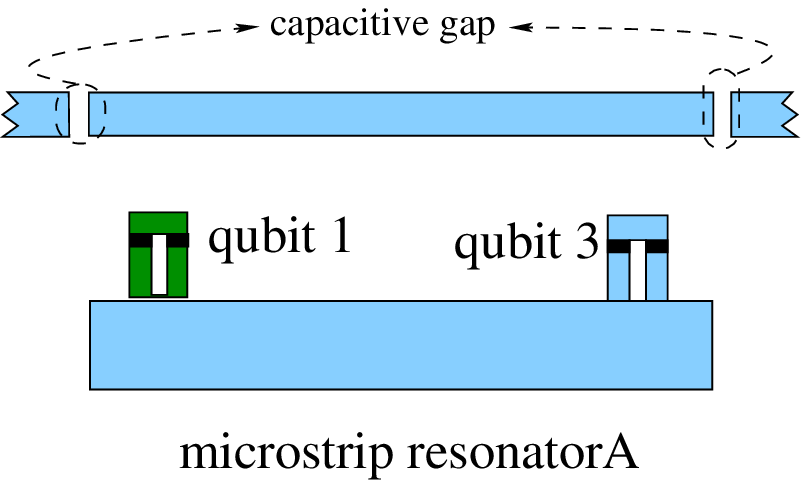,width=4.0cm,height=2.5cm}
\psfig{figure=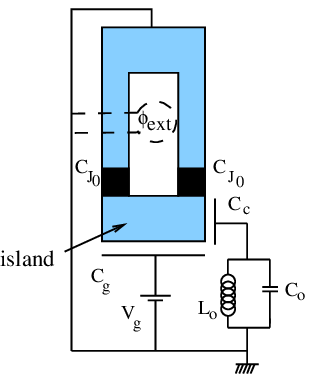,width=2.5cm,height=3.0cm}}
\caption{({\bf a}): Coupling between the field mode of
a microstrip resonator and two SQUID qubits. We show
just {\it one} resonator-qubits subsystem. This scheme
has to be doubled considering a resonator $B$ interacting
with qubits $2$ and $4$. Qubit $1$ and $2$ are used to
generate a MEMS, while the interaction of the cavity
modes with qubit $3$ and $4$ is useful to infer the
state of the field modes. ({\bf b}): Quantum circuit showing the LC-oscillator
model (parameters $L_{0},\,C_{0}$) for the cavity
mode. The coupling with the qubit is capacitive
(coupling capacitance $C_{c}$). $C_{J_{0}}$ is the
capacitance of each Josephson junction.}
\label{sistema2}
\end{figure}

A sketch of the set-up we consider is given in Fig.~\ref{sistema2} ({\bf a}).
Let us first concentrate on the interaction of 
a SQUID qubit with a single cavity mode. We operate at the  
SQUID degeneracy point where the qubit is encoded in equally-weighted superpositions of 
states having zero and one excess Cooper-pair on
the SQUID island, namely $\ket{\pm}=(1/\sqrt{2})\left(\ket{0}\pm\ket{2e}\right)$ 
($2e$ being the charge of a Cooper pair). The free Hamiltonian of the SQUID is given by
$H_{squid}=\frac{1}{2} E_{J}(\phi_{ext})\hat{\sigma}_{z}$. 
Here, $\hat{\sigma}_{z}$ is the $z$-Pauli operator
and $E_{J}(\phi_{ext})$ is the Josephson energy which is 
tuned via an external magnetic flux $\phi_{ext}$ piercing
the SQUID loop. This changes the energy separation between 
$\ket{+}$ and $\ket{-}$ and can be used to
switch on/off resonance the interaction with the field mode. 
The degeneracy point is set by biasing the SQUID with a
dc electric field via the ground plate of the resonator 
(Fig.~\ref{sistema2} ({\bf a})). To quantify the
strength of the qubit-field interaction, we model 
the cavity mode as a 
LC oscillator coupled to the SQUID as in
Fig.~\ref{sistema2} ({\bf b}). The coupling 
is realized via the 
capacitor $C_{c}$. The interaction Hamiltonian
can be cast in the form of a Jaynes-Cummings 
interaction with 
a Rabi frequency $\Omega$ that, for proper choices
of the circuital parameters, is as large as 
${0.5}\,GHz$~\cite{francescopino,mauro}. 
The {\it strong coupling regime}
is thus possible in this set-up and interaction times 
of about $ \sim 10 \,nsec\ll\kappa^{-1},\Gamma^{-1}$ 
are enough to generate an entangled state of two qubits near to the 
MEMS boundary.

Two SQUID qubits (size $\sim \mu m$) 
can be easily accommodated in the cavity ($L \sim 1 \, cm$) 
far enough to 
achieve negligible cross-talking (in principle due to 
direct capacitive and inductive coupling). 
Lithographic techniques allow to control (within few percent) the geometric characteristics and 
the resulting parameters of the device. 
The two qubits can be 
manipulated both simultaneously with a uniform magnetic field or independently with two separate coils. 
Due to charged impurities in the vicinity of the device,
separate calibration at the degeneracy points is 
required for each qubit. This may be achieved with several
adjustments of the design of ref.~\cite{schelkopf}, for
instance by splitting the ground plate and by attaching a gate to
each part.

The qubits have fixed position inside the cavities. This is 
an advantage with respect to common microwave cavity-QED systems, where flying atoms are required. The interaction
times are regulated tuning the interaction between the SQUID and the cavity mode on and off resonance via $\phi_{ext}$. This avoids the problem of different qubit injections in the cavities even if the two SQUIDs have to be set on
resonance with the respective cavity mode at the same time. The results we derived in Section~\ref{cavita}
about a mismatched interaction-triggering are fully applicable here. On the other hand, having precisely
fixed the number of qubits inside the cavities, we do not have to deal with spectator qubits.

We now briefly address the effect of decoherence in this set-up. 
Working at the charge degeneracy point, we minimize the 
effect of charge-coupled sources of noise~\cite{francescopino}. At this point, 
decoherence due to low-frequency modes vanishes at first order, a key property 
which allows to achieve coherence times of several hundreds of 
nanoseconds in the Saclay qubit~\cite{vion}. Furthermore, in our set-up,
 a selection rule prevents direct transitions
between the states of the dressed doublet~\cite{francescopino}. As a result of this analysis, the performance-limiting process turns out to be the spontaneous decay from each state of the dressed doublet to the ground state. This results
from comparable contributions of non-radiative decay of the 
qubit due to $1/f$ noise and losses due to the resonator. Life-times of $\sim 300\,ns$ were estimated for
a system where the resonator is implemented by a Niobium Josephson junction.
As we said, the losses of the resonator are further minimized in the proposal of 
ref.~\cite{schelkopf}, where $1/f$ noise is the ultimate
limiting factor. Noise sources are switching charged 
impurities and their effect depends on statistical properties of the 
environment which are beyond the power spectrum~\cite{elisabettalara}.
The actual dephasing rate depends also on details of 
the protocol and may show device-dependent 
features~\cite{oneoverf}. A detailed analysis with simulations
of noisy gates for this device will be presented 
elsewhere~\cite{falci}. It is worth stressing here that, due to the qubit-resonator interaction, the energy levels of our qubit are much less sensitive to charge fluctuations than an isolated qubit at the optimal working point. This leads to a conservative estimate of coherence times in the
range of $100\,ns$~\cite{falci}, which surely allows for navigation 
and generation of MEMS.

At the degeneracy point, computational states of the qubit cannot be distinguished by measuring their charge. 
To detect the state of a SQUID qubit, we have to slowly shift the working point far from the
degeneracy point, sweeping a dc-bias, to adiabatically
transform the states of the qubit as
$\ket{+}\rightarrow\ket{2e},\,\ket{-}\rightarrow\ket{0}$.
Then, charge measurements can be performed~\cite{Nakamura}.

As far as the interaction with more than one pair of qubits
is concerned, we make use of the further degree of freedom 
represented by the tunability of the energy spacing between the levels of each
qubit. We can proceed as follows. We suppose SQUID $1$
and $3$ in resonator $A$ (as in Fig.~\ref{sistema2} ({\bf a})) while SQUID $2$ and $4$ interact with $B$. First, just $1$
and $2$ are resonant with the corresponding cavity mode while $3$ and $4$ are in a dispersive regime obtained
either dc-biasing them or using an external magnetic flux. 
Once the interaction time $\tau_{1}$ has passed,
we set the interaction with this pair in a dispersive regime while $3$ and $4$ are set on resonance with the
respective cavity mode. The timing of these operation can be controlled electronically and the operating time can
be as large as $50\,nsec$. In this condition, an entangled state of $3$ and $4$ is established still within the
coherence time of the system.

In summary this second set-up offers some advantages 
with respect to a cavity-QED implementation. The most
important points are related to the longer coherence
times of the dynamical evolution while
this is the main limitation of an optical set-up.
An important problem for this solid state system is $1/f$ noise. 
In our opinion further improvements can 
be achieved by development of the design exploiting the 
possibilities of nanolithography.

\section{Conclusions}
When a quantum-correlated CV state, such as a two-mode squeezed state of light, interacts with a bipartite
two-dimensional system via bi-local interaction, effective entanglement transfer is possible. Theoretically, the
state of the qubits can be engineered in terms of entanglement and purity. The model provides a tunable source
of entangled mixed states that can be useful to investigate the interplay between entanglement and purity in
QIP and for purposes of quantum communication and computation. The entangler, {\it
i.e.} the CV system, does not exhaust its entangling capabilities with a single interaction but is able to
entangle other pairs of qubits. This property can be exploited as a test, based on the entangling power of the
field, for the quantum correlations in a non-Gaussian state of light.

We have proposed two different set-up in which this
scheme can be implemented. The first is a cavity-QED
system in which the qubits are embodied in two-level
atoms crossing two optical cavities. The second
proposal exploits the recent ideas about solid-state
systems/quantum optics interfaces
and uses superconducting qubits integrated in
microstrip resonators. This second scenario, in particular,
offers the advantages of a strong coupling regime of
interaction (that is hard to get with optical
cavities) without the difficulties connected with the
management of flying qubits.

\section*{Acknowledgments}
This work was supported by the UK Engineering and
Physical Science Research Council grant GR/S14023/01
and the Korea Research Foundation basic research grant
2003-070-C00024. MP and WS, respectively, thank the International Research Centre for Experimental Physics and the Overseas Research Student Award for financial support.



\begin{thebibliography}{99}
\bibitem{commentoM} This state is known to offer some advantages in terms of the amount of entanglement that can be transferred from the field modes to the qubits~\cite{mauro}.
\bibitem{commento} The spontaneous emission by the
qubits can be included considering the non-Hermitian term
$-i\Gamma\pro{e}{e}$ in
the energy of each qubit, $\Gamma$ being the
spontaneous emission
rate from $\ket{e}$. The effect of this term is a
decay of ${\cal E}_{NPT}$
with a rate dependent on $\Gamma$.
\end{thebibliography}
\end{document}